\documentclass[pdftex,prl,twocolumn,superscriptaddress,showpacs]{revtex4}

\usepackage{amsmath}
\usepackage{amssymb}
\usepackage[pdftex]{graphicx}
\usepackage{subfigure}

\begin{document}

\title{Anomalous Zeeman response in coexisting phase of superconductivity and spin-density wave as a probe of extended $s$-wave pairing structure in ferro-pnictide}

\author{Pouyan Ghaemi}
\affiliation{Department of Physics, University of California at Berkeley, Berkeley, CA 94720}
\affiliation{Materials Sciences Division, Lawrence Berkeley National Laboratory, Berkeley, CA 94720}

\author{Ashvin Vishwanath}
\affiliation{Department of Physics, University of California at Berkeley, Berkeley, CA 94720}
\affiliation{Materials Sciences Division, Lawrence Berkeley National Laboratory, Berkeley, CA 94720}

\date{\today}

 
\begin{abstract}
In several members of the ferro-pnictides,  spin density wave
 (SDW) order coexists with superconductivity over a range of dopings.
In this letter we study the anomalous magnetic Zeeman response of this coexistence state and show that it can be used to confirm the extended s-wave gap structure as well as structure of superconducting (SC) gap in coexisting phase.  On increasing the field, a strongly
anisotropic reduction of SC gap is found. The anisotropy is directly connected to the gap structure of superconducting phase. The signature of
this effect in quasiparticle interference measured by STM, as
 well as heat transport in magnetic field is discussed. For the compounds with the nodal SC gap  we show that the  nodes are removed upon formation of SDW. Interestingly the size of the generated gap in the originally nodal areas is anisotropic in the position of the nodes over the Fermi surface in direct connection with the form of SC pairing.
 \end{abstract}

 \maketitle The discovery of superconductivity at elevated temperatures in a new class of iron based materials (ferro-pnictides)\cite{iron}
has revived interest in the underlying mechanisms of high temperature superconductivity. Two key question are: what is the relation between antiferromagnetic order and superconductivity and what is the form of the superconducting pairing? These questions are fundamentally connected to one another. A remarkable feature of the ferro-pnictides is that in many cases they exhibit coexistence of superconductivity and antiferromagntism, over a wide range of parameters. As discussed below, studying the coexistence phase can reveal important information regarding both these questions.

A popular theoretical model for superconducting pairing is the extended $s$-wave ($s_\pm$)
pairing\cite{mazinx,ex1,pair2,ex2,fa,ex3,ex4,zlatko}, although other
pairing symmetries were also discussed \cite{pair1,pair3}.
In its simplest form, $s_\pm$ pairing consists of nodeless singlet
pairing but with different signs on different Fermi surface (FS)
pockets. Experimental studies on the nature of pairing have not yet
reached a unanimous conclusion. Hence new approaches to probing
pairing are desirable.


The parent compounds of many of the pnictides are antiferromagnetic
metals \cite{neutron}, referred to here as the spin density wave
(SDW) state.
On doping, the magnetic order is reduced, and superconductivity emerges. Establishing the detailed phase diagram, and whether these
two orders occur together, is an important question.
In some members of pnictide family e.g. $CeFeAsO_{1-x}F_x$\cite{nco}
SDW and SC phases have no overlap while in $LaFeAs(O,F)$
\cite{cco1,cco2} there are conflicting reports on SC and SDW phase
coexistence. On the other hand coexistence of SDW and
superconductivity in multiple pnictide materials have been reported.
For example, in $Ba_{1-x}K_xFe_2As_2$ and $Ba(Fe_{1-x}Co_x)_2As_2$
an extended region of coexistence with $0.2\leq x <0.4$ and
$0.025\leq x <0.06$ with a maximum superconducting transition
temperature inside this region of $\sim 28\ K$ and $20\ K$
respectively, are observed \cite{coex1,coex2}. Moreover, scanning
probe measurements of the latter indicate that the two orders
coexist in the same part of the sample\cite{KAM}.

\begin{figure}[htp]
\vspace{0.05in}\subfigure[]{\label{fsnd}
\includegraphics[height=4cm,width=4cm]{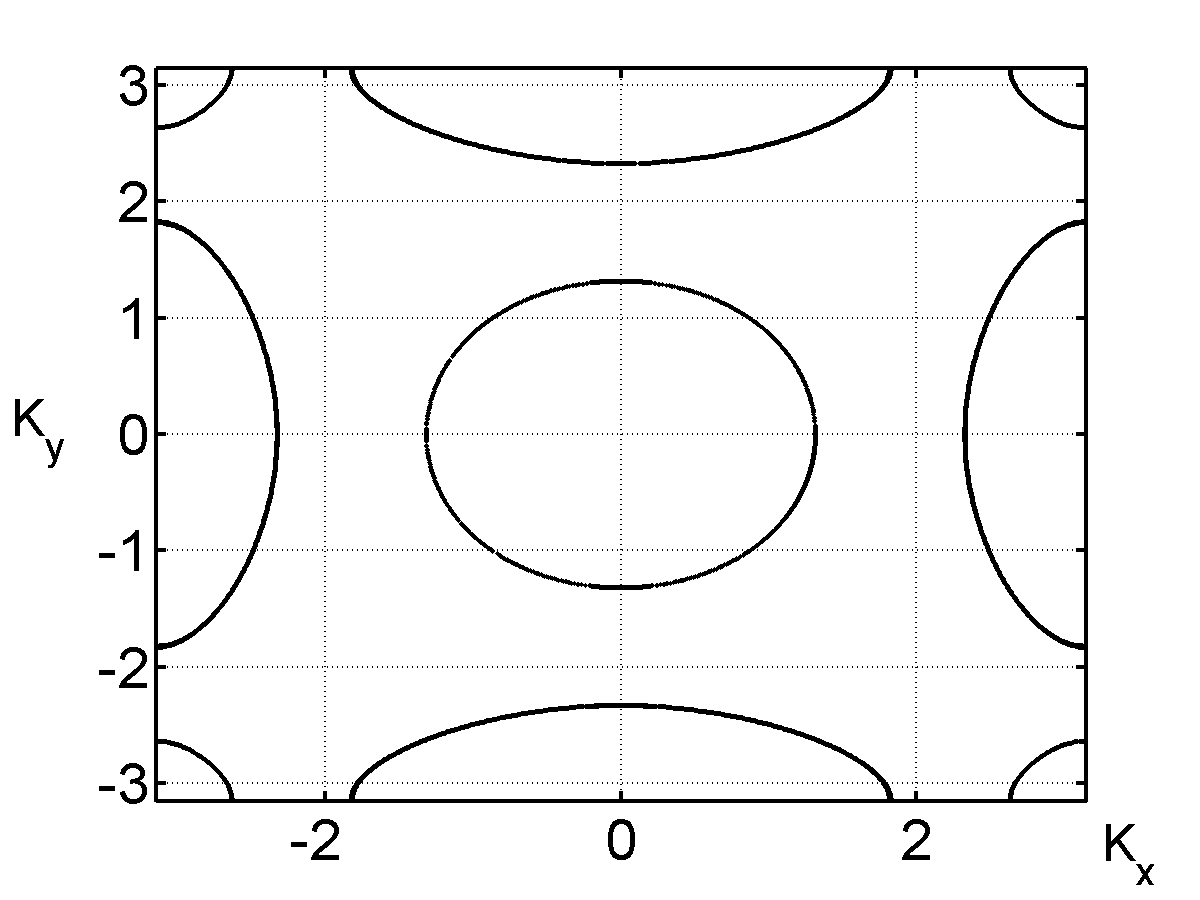} \vspace{0.1in}
} \subfigure[]{
\includegraphics[height=4cm,width=2cm]{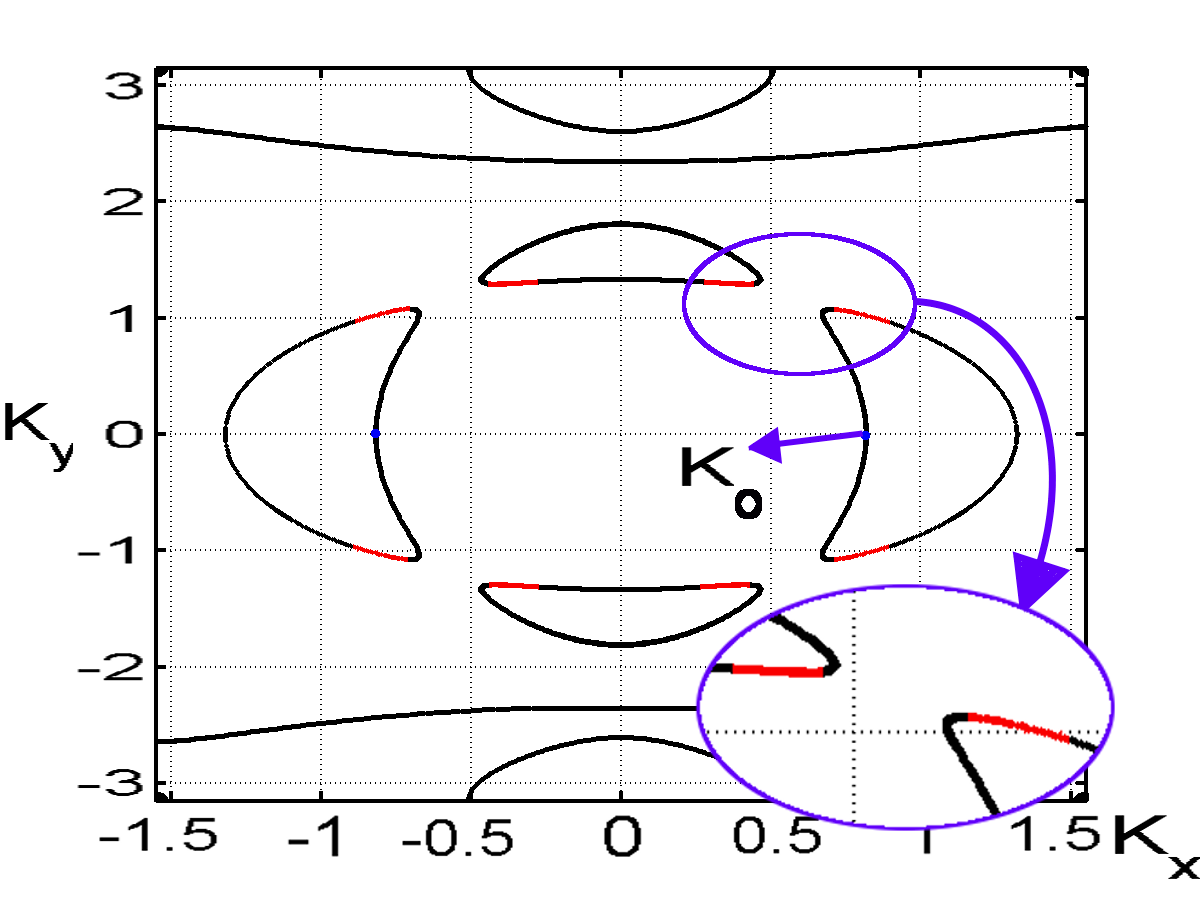}
\vspace{0.05in} \label{fssdw}}
\subfigure[]{
\includegraphics[height=2.3cm,width=4.9cm]{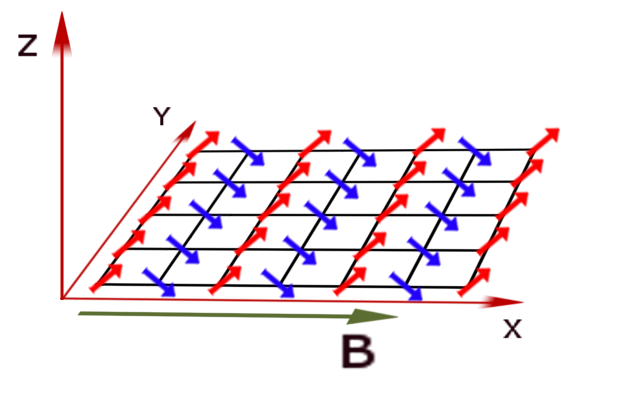}
\vspace{0.05in} \label{sys}}
\caption{a) FS in the first Brillouin zone. b) FS in the new Brillouin zone after SDW order forms. SC gap in red regions is robust against Zeeman. $\textbf{k}_0$ is the point where the gap closes first. c)SDW in transverse magnetic field}
\end{figure}

Theoretical studies of SDW coexisting with superconductivity has a
long history (see \cite{mazin} and references therein).
For the pnictides, with extended $s$-wave gap, the ordering wave vector
$\textbf{Q}=(\pi,0)$ \cite{neutron} folds the original Brillouin
zone (figure \ref{fsnd}) and parts of FS with opposite sign gap
cross.  It might seem that since  SC gap changes sign on the new FS
it has to have nodes. But the sign change of the gap upon
translation by $\textbf{Q}$
($\Delta^{\pm}(\textbf{q})\Delta^{\pm}(\textbf{q}+\textbf{Q})<0$) protects the nodeless SC gap \cite{mazin,Fernandes}. One simple way to
understand this is that in the clean limit, it is possible to
mathematically transform the  $s_\pm$ SC with SDW  problem, into an
$s$ wave SC with a charge density wave. Since the latter is an
s-wave SC with a time reversal symmetric perturbation, its gap is
protected by Anderson's theorem, and no nodes appear. 
This argument shows the possibility of coexistence of extended $s$-wave SC gap and SDW but it is not exclusive. Indeed as we show simple $s$-wave can also coexist with SDW despite the gap size reduces dramatically upon formation of SDW (also see \cite{chubnew}) so coexisting phase alone is not indication of form of SC pairing.   



In this letter we study an unusual aspect of the response of
the coexisting phase to a magnetic field which is direct evidence of extended $s$-wave gap structure. The orbital effect of the magnetic
field will be to produce vortices as in any type II superconductor.
Here, we will focus instead on the {\em Zeeman} part of the
coupling. In a layered superconductor which is appropriate for some of the
pnictide materials, a field applied in the plane has a weaker
orbital effect\cite{2D}. Moreover, the Zeeman coupling is effectively
enhanced by the presence of SDW order since those magnetic moments
will cant along an applied field. The electron $g$ factor is effectively increased 3 to 4 times over
its bare value. The transverse ferromagnetic moment that develops,
can decrease the SC gap. We show that this gap reduction occurs
in a highly nonuniform fashion.
The gap remains robust in some regions of FS (red regions in
fig. \ref{fssdw}) up to very high fields. In these
regions the pairing arises from the interplay of $s_\pm$ SC and SDW
orders and effectively has a triplet character. So for the first time we show that coexisting phase can exclusively test the extended $s$-wave SC gap structure. Consequences for STM experiments as well as possible signatures  of the Zeeman induced
gap closing over parts of the FS in heat transport are analyzed. 

In some members of pnictides family, the SC gap has nodes and it is suggested that nodes are result of superposition of simple and extended $s$-wave pairing\cite{fanodal}. For the first time we study the coexisting phase of such pairing structure with SDW and show that upon formation of SDW the nodes of this gap structure are removed with clear indication in multiple experimental probes e.g. heat capacity. The size of formed gaps are again strongly anisotropic in close connection with form of SC paring.

{\em Model:} The Hamiltonian including coupling of the mean field
SDW order with the conduction electrons is:

\begin{equation}\label{hamil}\begin{split}
\textbf{H}_{MF}= & \sum_{\textbf{k}} \Phi_{\textbf{k}}^\dagger \textbf{h}(\textbf{k})\Phi_{\textbf{k}}+H_{Spin} \\ & +\sum_i\left[\eta\left(e^{\textbf{Q}\cdot\textbf{r}_i}\ \textbf{S}_{AF}+\textbf{S}_F\right)+\frac{g}{2} \mu_B \textbf{H}\right]\cdot\Phi_i^\dagger \vec{\sigma} \Phi_i
\end{split}\end{equation}

Here $\Phi^T=\left( \psi_\uparrow, \psi_\downarrow \right)$ and $\psi_s$ is the orbital spinor. For simplicity, in this letter we use a two band
model for the pnictides\cite{ying,raghu} which already captures the
essential physics. So $\psi_s^T=\left( d_{xz}, d_{yz} \right)_s$ is the two component spinor
corresponding to two $d$ orbitals at the Fermi energy and the
kinetic Hamiltonian is given by\cite{ying}:

\begin{align}
&\textbf{h}_k(\textbf{k})= 2 t_1(\cos k_x-\cos k_y)\lambda^z + 2(t_2-t_2')\sin k_x \sin k_y\lambda^x \notag\\
& +[ 2(t_2+t_2')\cos k_x\cos k_y+2 t_1'(\cos k_x+\cos
k_y)-\mu_f]\cdot \lambda^0
\end{align}
where $t_1$($t'_1$) and $t_2$($t'_2$) represent nearest and next nearest neighbor hoping parameters, $\mu_f$ is the chemical potential and
$\lambda$ is the Pauli matrix acting on the orbital space.

$H_{spin}$ corresponds to the spin interaction which leads to
formation of $\textbf{Q}=(\pi,0)$ ordering (e.g. $J_1-J_2$
model\cite{magneticth,epl}). Last term in (\ref{hamil}) corresponds to
coupling of ordered moments and the Zeeman field ($\textbf{H}$) with conduction electrons ($\sigma$ is
the Pauli matrix acting on the Physical spin space).
$\textbf{S}_{AF}$ and $\textbf{S}_F$ are SDW moment and
ferromagnetic moment generated by transverse field.

{\em Zeeman Coupling Strength:} We now estimate the magnitude of the
transverse Zeeman coupling ($\Delta_{Ferro}=
|\eta\textbf{S}_{F}+\frac{g}{2}\mu_B\textbf{H}|$). Note, the SDW
will flop into the plane perpendicular to the field, for
sufficiently weak magnetic anisotropy, so the field is transverse to
the SDW moment. To estimate the ferromagnetic moment
$\textbf{S}_{F}=\chi \textbf{H}$, we use the transverse
susceptibility in the non-SC SDW phase of the $1111$ compounds:
$\chi\sim 0.4\times 10^{-4}\ emu$ \cite{epl}. Since $\eta$ factor in
(\ref{hamil}) is not known, we can not estimate $\Delta_{Ferro}$
directly. The ferromagnetic moment and SDW moment couple to the
conduction electrons similarly so we can use the properties of SDW
phase to estimate $\Delta_{Ferro}$. The magnitude of SDW moment is $S_{AF} \sim 0.36\ \mu_B$\cite{neutron} and the SDW
gap $\Delta_{SDW}=|\eta\textbf{S}_{AF}| \sim 0.08\ eV$\cite{sdwgap}
is measured using optical spectroscopy. 
The effective change in electron $g$ factor is then $\delta
g=\frac{2 \ \chi}{|\textbf{S}_{AF}|}\frac{\Delta_{SDW}}{\mu_B}\sim
5.5$. The SC gap $\Delta_s \sim 4\ meV$ is measured using ARPES\cite{arpes}. In a transverse field of $\textbf{H} \sim 18 T$, $\Delta_{Ferro} \sim \Delta_s$. In the layered pnictides, a critical magnetic field
of order $50 T$ have been reported\cite{2D} so Zeeman coupling
effect might indeed be as relevant as orbital effects in these
fields.

{\em Coexistence Phase:} To study the coexisting phase we use an extended Hilbert space and
consider states at \textbf{k} and $\textbf{k}+\textbf{Q}$ as two
component pseudospin: $ \Psi_{\textbf{k}}^T= \left(
\Phi_{\textbf{k}}, \Phi_{\textbf{k+Q}}\right) $. The Hamiltonian (\ref{hamil}) with $\textbf{S}_{AF} || \hat{z}$ and $\textbf{S}_F || \hat{x}$ (fig. \ref{sys})
will be:

\begin{equation}\label{khamil}\begin{split}
\textbf{H}_{MF}(\textbf{k})=& \left(\begin{array}{ccc} \textbf{h}_k(\textbf{k}) & 0 \\ 0 & \textbf{h}_k(\textbf{k+Q})\end{array}\right) \sigma^0 \\ & +\Delta_{SDW} \lambda^0\tau^x \sigma^z+\Delta_{Ferro} \lambda^0\tau^0\sigma^x
\end{split}\end{equation}

 $\tau$ is the Pauli matrix acting on $(\textbf{q},\textbf{q}+\textbf{Q})^T$ space. Diagonalizing $\textbf{h}_k(\textbf{k})$ we get the two bands. At each $\textbf{k}$ point we project in to the state closer to the Fermi energy which is a two component wave function $\psi_s(\textbf{k})$. Then $\textbf{h}_k(\textbf{k})$ will be replaced by the corresponding eigenvalue $\epsilon(\textbf{k}) \lambda^0$. Projecting the Hamiltonian (\ref{khamil}) into the low energy orbital space we get:

\begin{equation}\begin{split}
\textbf{H}_{MF}^P(\textbf{k})= & \left[ E^+(\textbf{k})\tau^0+E^-(\textbf{k})\tau^z \right] \lambda^0 \sigma^0\\ &+\Delta_{SDW} \lambda^0\tau^x \sigma^z+\Delta_{Ferro} \lambda^0\tau^0\sigma^x
\end{split}\end{equation}

where $E^\pm(\textbf{k})=\frac{\epsilon(\textbf{k})\pm\epsilon(\textbf{k}+\textbf{Q})}{2}$. When $\Delta_{Ferro}=0$ the dispersion is $\varepsilon(\textbf{k})=E^+(\textbf{k})\pm\sqrt{E^{-2}(\textbf{k})+\Delta_{SDW}^2}$. Assuming that the SDW ordering does not change the chemical potential, for the points on the FS we have $\varepsilon(\textbf{k})=0 \Rightarrow E^{+2}(\textbf{k})=E^{-2}(\textbf{k})+\Delta_{SDW}^2$.

The SC Hamiltonian acting on $\left(\Psi_{\textbf{k}}, \Psi_{-\textbf{k}}^\dagger\right)^T$ is: 

\begin{equation}\label{shamil}\begin{split}
& \textbf{H}(\textbf{k})= E^+(\textbf{k})\tau^0\sigma^0\mu^z +  E^-(\textbf{k}) \tau^z\sigma^0\mu^z+\Delta^{\pm}_{sc}(\textbf{k}) \tau^z\sigma^y\mu^y\\ &  +\Delta^0_{sc}(\textbf{k}) \tau^0\sigma^y\mu^y + \Delta_{SDW}\ f(\textbf{k})\  \tau^x\sigma^z\mu^z+\Delta_{Ferro} \tau^0\sigma^x\mu^z
\end{split}\end{equation}

where $\mu$ is the Pauli matrix acting on SC particle-hole space. $\Delta^{0}_{SC}$ is the the ``simple" $s$-wave ($\Delta^{0}_{SC}(\textbf{k})=\Delta^{0}_{SC}(\textbf{k+Q})$) and $\Delta^{\pm}_{SC}$ is the extended $s$-wave ($\Delta^{\pm}_{SC}(\textbf{k})=-\Delta^{\pm}_{SC}(\textbf{k+Q})$) part of pairing. We have traced over the orbital ($\psi_s$) space leading to $f(\textbf{k})=\sum_s \psi_s^\dagger(\textbf{k}) \psi_s(\textbf{k}+\textbf{Q})$ in the $SDW$ term which mixes the states at $\textbf{k}$ and $\textbf{k}+\textbf{Q}$. This term is important in the perfect nesting limit and is ignored otherwise.


First we consider $\Delta_{Ferro}=0$ where the Hamiltonian could be diagonalized analytically. Defining $A^2(\textbf{k})=2E^+(\textbf{k})^2+\Delta_{sc}^{\pm^2}+\Delta^{0^2}_{sc}$, $E^2(\textbf{k})=A^2(\textbf{k})-2\sqrt{\frac{A(\textbf{k})^4-(\Delta_{sc}^2-\Delta_{sc}^{02})^2}{4}-\left(E^+(\textbf{k})\Delta_{sc}-E^-(\textbf{k})\Delta_{sc}^0\right)^2}$. Many features of coexisting phase could be understood from this dispersion: extended and ``simple" s-wave pairing can both coexist with SDW although the gap for ``simple" s-wave reduces greatly upon formation of SDW. If the SC gap has nodes as a result of presence of both ``simple" and extended s-wave paring\cite{fanodal}, upon formation of SDW the nodes are removed, i.e. magnetism enhances the SC properties! In general the gap formed is small \cite{chubnew} but is enhanced if nodes happen to occur close to the nesting regions. In the rest of paper $\Delta_{sc}^0=0$ and use $\Delta_{sc}=\Delta^{\pm}_{sc}$.

{\em Effect of Magnetic Field:} Although it seems that all the symmetries of the Hamiltonian (\ref{shamil}) are broken by SC and SDW orders, there is a remaining symmetry implemented by the operator $\Sigma = \tau^z\sigma^x\mu^z$ which commutes with the Hamiltonian in (\ref{shamil}). $\Sigma$ has four eigenstates with eigenvalue $1$ and four with eigenvalue $-1$. We can reduce the size of the Hamiltonian in (\ref{shamil}) by projecting into the subspaces corresponding to different eigenvalues of $\Sigma$. 

For zero energy eigenvalue in the characterizing polynomial we can see that after the ferromagnetic moment is formed, at each $\textbf{k}$ point the SC gap vanishes when:
\begin{equation}\label{gc}
\Delta_{Ferro}^2=2 E^-(\textbf{k})^2\left(1-\sqrt{1-\frac{\Delta_{SDW}^2 \Delta_{SC}^2}{E^-(\textbf{k})^4}}\right) +\Delta_{sc}^2
\end{equation}

Here one can readily see that the gap vanishes anisotropically, since $E^-(\textbf{k})$ varies over FS, even though we consider magnitude of $\Delta_{SC}$ to be uniform. At the point where the SC gap vanishes first, the low energy excitations dispersion is anisotropic: $\varepsilon(\textbf{p})=\frac{\Delta_{SDW}^2\Delta_{SC}}{E^-(\textbf{k}_0)^3}\left(\alpha |p_x|+\beta p_y^2\right)$. $\alpha$ and $\beta$ depend on the the band curvatures, $\textbf{k}_0$ is the position where the gap first closes (fig. \ref{fssdw}) and $\textbf{p}$ denotes deviation from $\textbf{k}_0$. On the other hand (\ref{gc}) shows that when  $|\Delta_{SDW} \Delta_{SC}|> E^-(\textbf{k})^2$ ferromagnetic moment can not close the gap! More specifically one can look at the point where $E^-(\textbf{k})=0$. Eigenvalues at this point could be calculated exactly. Defining $\ell^2=2\sqrt{\Delta_{SDW}^2\Delta_{Ferro}^2+\Delta_{SC}^2\Delta_{Ferro}^2+\Delta_{SDW}^4}$, the energy is
$E^2=\sqrt{\ell^4+(\Delta_{Ferro}^2-\Delta_{SC}^2)^2+4\Delta_{SC}\Delta_{SDW}}\pm\ell^2>0$.

The eigenvalues are non-zero as long as both SDW and singlet superconductivity are present, regardless of the ferromagnetic moment. It is important to note that the gap is not SDW gap but it is indeed a SC gap (it vanishes when $\Delta_{SC}=0$) that is robust against external magnetic field; as we will show it is indeed spin-triplet pairing gap.

Two aspects of these results are particularly puzzling; since the singlet pairing changes sign on the FS, it seems that it should vanish at some points, but our result indicates that nodeless superconductivity  coexist with SDW. The other feature is that the coupling with ferromagnetic moment  affects the SC gap anisotropically and it can not destroy the gap in some regions. Below we show that the nature of the SC gap can explain these puzzling features.

When $\Delta_{Ferro}=0$ eigenvalues and eigenstates of the Hamiltonian for the points on the FS could be derived analytically. Interestingly two operators corresponding to the spin-triplet pairing also commute with the $\Sigma$. Operator that is important for us is $\Gamma_{Triplet}=\tau^y\sigma^x\mu^y$ which anticommutes with the singlet operator $\Gamma_{Singlet}=\Delta_{SC}\tau^0\sigma^y\mu^y$. The amplitude for any type of pairing could be calculated self-consistently using the wave functions $|\Psi_n\rangle$ \cite{scbook} as $\Delta_p = \Delta_{SC}\sum_{n, E_n<0} \langle \Psi_n | \Gamma_p | \Psi_n \rangle$:

\begin{eqnarray}
& \Delta_S(\textbf{k})& \propto  \sum_n \langle \Psi_n | \Gamma_{Singlet} | \Psi_n \rangle \\ =&\Delta_{SC}& \frac{E^-(\textbf{k})\left(\sqrt{E^-(\textbf{k})^2+\Delta_{SDW}^2}-E^-(\textbf{k})\right)}{\Delta_{SDW}^2-E^-(\textbf{k})\left(\sqrt{E^-(\textbf{k})^2+\Delta_{SDW}^2}-E^-(\textbf{k})\right)}  \notag \\
& \Delta_T(\textbf{k})& \propto  \sum_n \langle \Psi_n | \Gamma_{Triplet} | \Psi_n \rangle \\
=&-\Delta_{SC}& \frac{\Delta_{SDW}}{\sqrt{\Delta_{SDW}^2+E^-(\textbf{k})^2}} \notag \\
&\Delta_S(\textbf{k})^2&+\Delta_T(\textbf{k})^2=\Delta_{SC}^2\label{sqconst}
\end{eqnarray}

As we expected $\Delta_S(\textbf{k})$ vanishes where
$E^-(\textbf{k})=0$ (it is where singlet pairing changes sign).
Around this point
$\Delta_S(\textbf{k})\approx\Delta_{SC}\frac{E^-(\textbf{k})}{\Delta_{SDW}}$
so it satisfies the expectation that singlet pairing should change
sign between the regions coming from different FSs after zone
folding.  On the other hand when $E^-(\textbf{k})=0$,
$\Delta_T(\textbf{k})\approx\Delta_{SC}$ so the pairing is triplet
type. In the opposite limit $E^-(\textbf{k}) \gg \Delta_{SDW}$,
$\Delta_S(\textbf{k})\approx\Delta_{SC}$ and
$\Delta_T(\textbf{k})\approx\Delta_{SC}
\frac{\Delta_{SDW}}{|E^-(\textbf{k})|}$ so the pairing is mainly
singlet. This even parity triplet pairing is robust against coupling
with the ferromagnetic moment i.e. regions with large triplet
pairing remain gapped as the ferromagnetic moment forms. Eqn.
(\ref{sqconst}) also shows that singlet and triplet pairings together
gap out all of the FS.

A special limit ``perfect nesting"
(with $t'_1=0$)\cite{ying}. In this limit $E^+(\textbf{k})=0$. It
might seems that SDW gaps out all parts of the FS. But here
 $f(\textbf{k})$ plays an important role as it vanishes linearly at
symmetry protected points on the FS\cite{ying}.
With superconductivity a full gap opens, which closes at
$\textbf{k}_0$ on increasing the field when $|\Delta_{Ferro}
|=|\Delta_{SC}|$.
The dispersion then is `semi-Dirac' like \cite{SinghPickett}
$E(\textbf{k}_0+\textbf{p})=\sqrt{v_F^2p_x^2+\rho p_y^4}$. We do not
discuss this case further since it requires fine-tuning to reach.

{\em Experimental Consequences:} So far we have proposed a
theoretical picture to understand the coexistence of
SC and SDW phase in ferro-pnictides. We showed
transverse Zeeman field reduces the SC gap anisotropically.
In the rest of this letter we will discuss the experimental
signature of the effect discussed above.
The usual experimental tool to map out the dispersion is ARPES which is not suitable in magnetic
field. Another approach which by now is widely used to map the FS (e.g. in cuprate superconductors) is the
quasiparticle interferences measurement using
STM\cite{stm1,stm2}. STM measures the local density of state which
is uniform for a normal clean metal. When sources of disorder such
as impurities or crystal defects are present, elastic scattering
mixes eigenstates that have different momentum but are located
on the same contour of constant energy (STM bias voltage). When scattering mixes states
$\textbf{k}_1$ and $\textbf{k}_2$, an interference pattern with wave
vector $\textbf{q}=\textbf{k}_1-\textbf{k}_2$ appears in local
density of states modulations which could be observed by STM as
modulations of differential tunneling conductance\cite{stm3}. The amplitude of the oscillation at momentum $\textbf{q}$ is proportional to the joined density of state at momentum $\textbf{q}$ i.e. $n(\textbf{k}+\textbf{q})n(\textbf{q})$ where $n(\textbf{k})$ is the density of state at energy equal to the STM bias voltage and momentum $\textbf{k}$.

\begin{figure}[htp]
\vspace{0.07in}\subfigure[]{\label{conste}
\includegraphics[height=2.8cm]{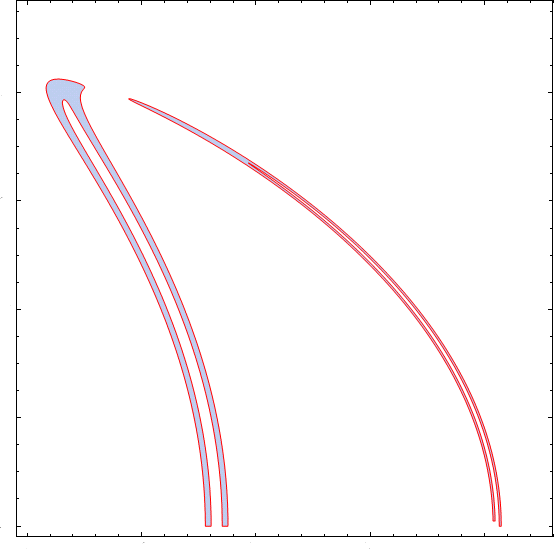} \vspace{0.1in}
} \subfigure[]{
\includegraphics[height=4.6cm]{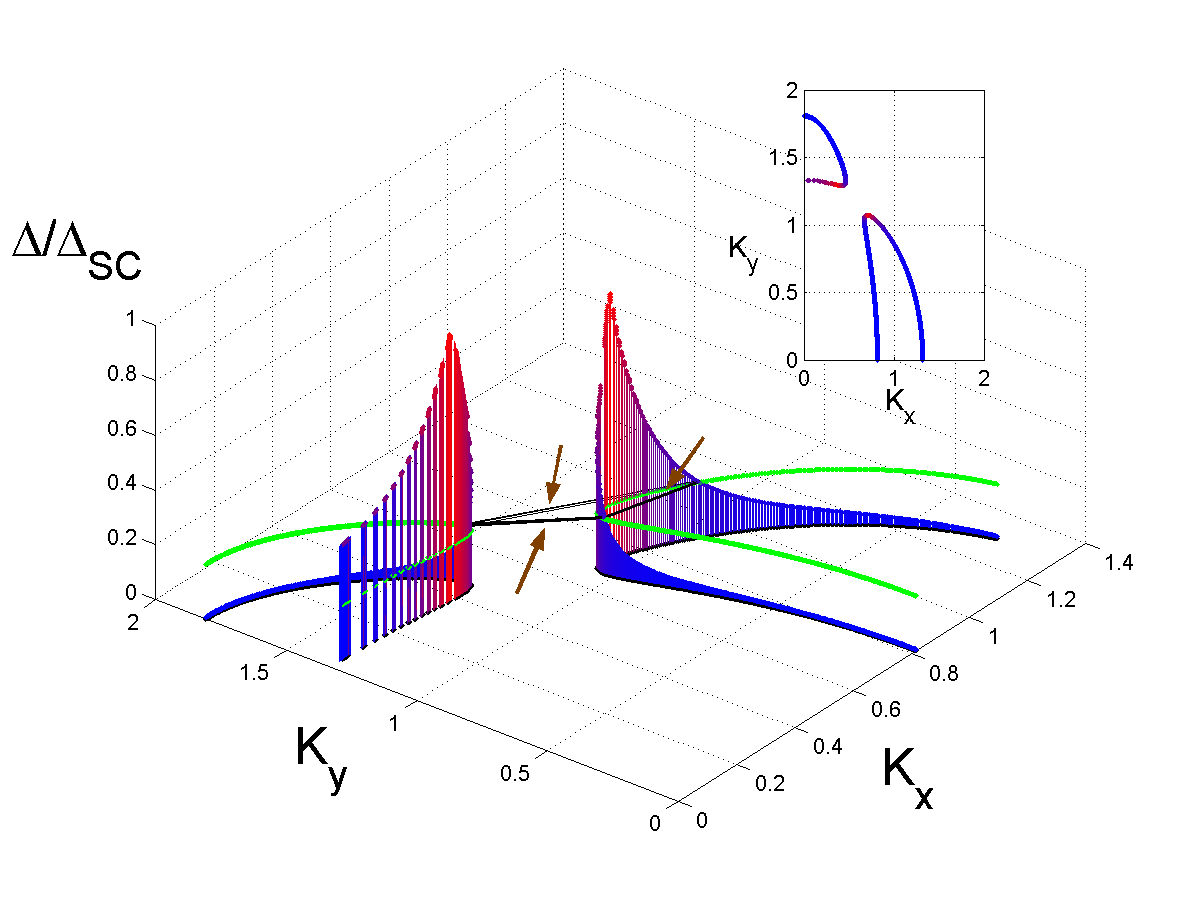}
\vspace{0.1in} \label{gap}}
 \caption{ a) Parts of constant energy contours. The thickness of blue region corresponds to the density of states. b)Variation of SC gap over parts of the FS in transverse magnetic field. Red and blue colors correspond the magnitude of triplet and singlet pairing respectively. The lines marked by arrows connect regions with large density of states at the energy marked by green curve. Fourier transform of density modulations have enhanced amplitudes at these momentums.}
\end{figure}

 The Zeeman field will generate
the variation of the superconducting gap over the FS. The dispersion
along the original FS is very shallow (at list of order
$\Delta_{SDW}\Delta_{SC}$) compared to direction perpendicular to
the FS. So the density of states at the tips of
constant energy curves (fig. \ref{conste}) is larger. The joint density of state for momentums connecting these regions (marked by arrows in figure \ref{gap}) is increased.
 These momentums
 vary continuously as STM bias voltage
changes and will give the complete map of dispersion relation. Without external field, there is no gapless excitation. As external
field is turned on the gap start to reduce anisotropically: vanishes in some regions but is not affected where
$E^-(\textbf{k})^2\ll\Delta_{SDW}\Delta_{SC}$.

Recent results on low temperature thermal conductivity of pnictide
superconductors \cite{Taillefer1,Taillefer2} indicate the presence
of a full gap over the FS which is highly sensitive to the external
magnetic field. The mechanism presented in this paper also leads to
partial destruction of the SC gap in magnetic fields much smaller
than the critical field. Based on our estimate using a uniform gap
magnitude, this field is still much larger than the range where
experiments have been performed. However, an anisotropic gap (as
found in some calculations \cite{fa}) could lead to much smaller onset fields
where the ferromagnetic moment generates nodes in the SC gap. When
the gap first vanishes as external field increases, the low energy
excitations have anisotropic dispersion
$\varepsilon(\textbf{p})\propto \alpha |p_x|+\beta p_y^2$
which leads to density of states $N(E)\propto \sqrt{E}$. The
signature of such density of states will be seen in temperature
dependence of superfluid density $\Delta \rho_s \propto -\sqrt{T}$,
as well as field dependence of heat conductivity $\Delta\kappa
\propto H^{\frac{1}{4}}$ (this could be understood as the Doppler
shift due to superfluid flow around the vortex\cite{volovik}). Note,
in contrast a Dirac node dispersion would have $\Delta \rho_s
\propto -T$ and $\Delta\kappa \propto H^{\frac{1}{2}}$.

{\em Conclusion:} We considered the effect of Zeeman coupling in
suppressing the extended $s$-wave SC gap in the phase where
superconductivity and SDW coexist and the transverse susceptibility
is enhanced. We showed that a highly anisotropic suppression of SC
gap in Zeeman field is indeed directly related to the extended
$s$-wave structure of SC gap. We acknowledge insightful discussions
with Fa Wang and support from LBNL DOE-504108.

\bibliography{coexist6}
\end{document}